**Improper ferroelectricity in stuffed aluminate sodalites for pyroelectric energy harvesting**

Yusaku Maeda[1], Toru Wakamatsu[1], Ayako Konishi[2,3], Hiroki Moriwake[2,3], Chikako Moriyoshi[4], Yoshihiro Kuroiwa[4], Kenji Tanabe[1], Ichiro Terasaki[1], and Hiroki Taniguchi[1,*]

[1]Department of Physics, Nagoya University, Nagoya 464-8602, Japan.

[2]Nanostructures Research Laboratory, Japan Fine Ceramics Center, Nagoya 456-8587, Japan.

[3]Center for Materials Research by Information Integration (CMI[2]) National Institute for Materials Science (NIMS), Tsukuba, 305-0047, Japan

[4]Department of Physical Science, Hiroshima University, Higashihiroshima, 739-8526, Japan.

[*]Corresponding author. hiroki_taniguchi@cc.nagoya-u.ac.jp

**ABSTRACT**

Ferroelectricity in stuffed aluminate sodalites, $(Ca_{1-x}Sr_x)_8[AlO_2]_{12}(WO_4)_2$ ($x \leq 0.2$) ($C_{1-x}S_xAW$), is demonstrated in the present study. Pyroelectric measurements clarified switchable spontaneous polarization in polycrystalline $C_{1-x}S_xAW$, whose polarization values were on the order of $10^{-2}$ μC/cm$^2$ at room temperature. A weak anomaly in the dielectric permittivity at temperatures near the ferroelectric transition temperature suggested improper ferroelectricity of $C_{1-x}S_xAW$ for all investigated values of $x$. A comprehensive study involving synchrotron X-ray powder diffraction measurements, molecular dynamics simulations, and first-principles calculations clarified that the ferroelectric phase transition of $C_{1-x}S_xAW$ is driven by freezing of the fluctuations of WO$_4$ tetrahedra in the voids of an $[AlO_2]_{12}^{12-}$ framework. The voltage response and electromechanical coupling factor of $C_{1-x}S_xAW$ estimated from the present results indicate that this material exhibits excellent performance as a pyroelectric energy harvester, suggesting that aluminate sodalites exhibit great promise as a class of materials for highly efficient energy harvesting devices.

## I. INTRODUCTION

Inducing a polar ground state is an intriguing issue in solid-state physics, materials science, and inorganic chemistry because the polar symmetry provides a unique background for various functionalities [1]. One representative example is ferroelectricity, which features switchable spontaneous polarization, pyroelectricity, piezoelectricity, and nonlinear optical responses and is applied in state-of-the-art technologies. Ferroelectric oxides have conventionally been developed with oxygen-octahedra-based compounds typified by perovskite-type ferroelectrics such as $Pb(Zr,Ti)O_3$, $BaTiO_3$, and $LiNbO_3$. Several effects play important roles in inducing ferroelectricity in octahedra-based oxides: A second-order Jahn–Teller effect gives rise to polar displacements of cations in the oxygen octahedra [2–8]. A steric hindrance effect of specific elements with a lone electron pair, by contrast, displaces the atom from the centric position among surrounding ligand oxygens to render the crystal noncentrosymmetric [9,10]. Furthermore, recent studies have suggested that multiple instabilities of octahedral rotations cooperatively induce ferroelectricity [11–15].

In contrast to the numerous investigations of oxygen-octahedra-based ferroelectric compounds, the literature contains few reports of ferroelectricity in oxygen-tetrahedra-based compounds other than hydrogen-bonded compounds [16], which are often unsuitable for practical applications because they deliquesce. Because oxygen-tetrahedra-based compounds are abundant in the Earth's crust, as typified by quartz and zeolites, the development of ferroelectric oxygen-tetrahedra-based compounds should lead to innovative eco-friendly devices. Recently, Taniguchi *et al*. reported excellent ferroelectricity in $Bi_2SiO_5$, an oxygen-tetrahedra-based silicate

[17]. The ferroelectricity in $Bi_2SiO_5$ is driven by twisting deformations of one-dimensional chains of $SiO_4$ tetrahedra that partially compose the crystal structure [17–19]. Furthermore, the switchable spontaneous polarization of approximately 23 $\mu C/cm^2$, which is comparable to that of perovskite-type ferroelectric oxides, is induced by internal deformation of $SiO_4$ tetrahedra [17,18,20]. The discovery of ferroelectricity in $Bi_2SiO_5$ has opened a path for the development of ferroelectric oxides through exploitation of oxygen-tetrahedra networks.

An aluminate sodalite-type oxide is an oxygen-tetrahedra-based oxide of the sodalite family [21]. It possesses an $[AlO_2]_{12}^{12-}$ framework structure composed of $AlO_4$ tetrahedra. The $[AlO_2]_{12}^{12-}$ framework has two inequivalent voids occupied by alkaline-earth divalent cations and $(MO_4)^{2-}$ ($M$ = S, Cr, Mo, W) tetrahedral anions, resulting in a chemical composition of $Ae_8[AlO_2]_{12}(MO_4)_2$ ($Ae$ denotes an alkaline-earth element). The aluminate sodalite-type oxides undergo various structural phase transitions because of the underlying lattice instability in their network of oxygen tetrahedra [22–24]. $Sr_8[AlO_2]_{12}(CrO_4)_2$, in particular, has been reported to undergo a ferroelectric phase transition at $T_c$ approximately 300 K [25]. A small spontaneous polarization and a weak dielectric anomaly observed at approximately $T_c$ indicate an improper ferroelectric phase transition of $Sr_8[AlO_2]_{12}(CrO_4)_2$, whose primary order parameter is not polarization but another physical quantity. Such improper ferroelectricity has recently been attracting increasing attention because of its potential for multi-functionalities based on intimate coupling between the polarization and another physical quantity, as observed in the case of multiferroic materials [26,27]. The aluminate sodalite-type oxides represent a new frontier in the development of oxygen-tetrahedra-based ferroelectric oxides.

Here, we report improper ferroelectricity in $(Ca_{1-x}Sr_x)_8[AlO_2]_{12}(WO_4)_2$ ($0 \leq x \leq 0.2$), which is abbreviated hereafter as $C_{1-x}S_xAW$. Pyroelectric measurements demonstrate switchable spontaneous polarization on the order of $10^{-2}$ $\mu C/cm^2$, which is comparable to that of conventional improper ferroelectric oxides [28]. Molecular dynamics simulations indicate that strong fluctuations of WO$_4$ tetrahedra in the voids of the [AlO$_2$]$_{12}^{12-}$ framework freeze to drive the improper ferroelectric phase transition. First-principles calculations clarify that the fluctuation of WO$_4$ tetrahedra stems from multiple instabilities at finite wavenumbers in acoustic phonon branches. Although its magnitude of spontaneous polarization is small, C$_{1-x}$S$_x$AW is found to possess strong potential for use in pyroelectric energy harvesters because of its small dielectric permittivity even at temperatures near the ferroelectric phase-transition temperature. The present study sheds new light on the development of oxygen-tetrahedra-based ferroelectric materials with potential applications in high-performance energy harvesting devices.

## II. EXPERIMENTAL

To synthesize polycrystalline samples of C$_{1-x}$S$_x$AW, stoichiometric mixtures of CaCO$_3$, SrCO$_3$, WO$_3$, and Al$_2$O$_3$ were heated at 1673 K for 12 h, followed by furnace cooling. Powder X-ray diffraction (PXRD) measurements confirmed that all samples were in a single-phase state with no impurity phases. Differential scanning calorimetry (DSC) was carried out on a DSC 8230 (Thermo Plus Co.) to investigate the phase-transition temperature. Dielectric permittivity was measured from room temperature to 700 K using a Keysight 4284A precision LCR-meter. The temperature dependence of the spontaneous polarization was obtained through measurement of the pyrocurrent during zero-field heating after poling by field-cooling using a bias field of 10 kV/cm. PXRD analyses were performed on the powder diffraction beamline BL02B2 at SPring-8 with the approval of the Japan Synchrotron Radiation Research Institute (JASRI) (proposal numbers 2014B1468 and 2015A1425) of SPring-8. The wavelengths of incident X-rays used in the present study were 18 keV (0.67 Å) and 30 keV (0.41 Å).

First-principles calculations were performed within the framework of density functional theory (DFT) using the projector-augmented wave (PAW) method [29] as implemented in the VASP code [30]. A plane-wave cutoff energy of 500 eV and a $4 \times 4 \times 4$ $k$-points mesh were used. We also carried out systematic first-principles molecular dynamics (FPMD) simulations within the NVT ensemble using the Nosé thermostat [31]. The time step was 1 fs. Simulation cells contained 54 atoms (i.e., 1 formula unit). The lattice volume optimized by DFT calculations for the high-temperature $I\bar{4}3m$ phase was used for all of the FPMD calculations.

## III. RESULTS AND DISCUSSION

The $C_{1-x}S_xAW$ is composed of an $[AlO_2]_{12}^{12-}$ framework with two nonequivalent voids filled by $Ca^{2+}(Sr^{2+})$ and $WO_4^{2-}$ (Fig. 1(a)). As shown in Fig. 1(b), one of the end members, $C_{1-x}S_xAW$ ($x = 0$), undergoes a successive high-temperature structural phase transition from a centrosymmetric $I\bar{4}3m$ phase (α-phase) to a noncentrosymmetric $Aba2$ phase (γ-phase) through an intermediate incommensurate phase (β-phase), which has been previously reported to exist in a narrow temperature range between the $I\bar{4}3m$ and $Aba2$ phases [32]. The other end member, $C_{1-x}S_xAW$ ($x = 1$), by contrast, has a high-temperature $Im\bar{3}m$ phase and the structure transforms to $I4_1/acd$ at approximately 600 K with decreasing temperature [33]. Figures. 1(b) and 1(c) show the phase diagram of $C_{1-x}S_xAW$ for $0 \leq x \leq 0.20$, as constructed on the basis of the results of DSC and PXRD analyses. Solid circles, triangles, and squares are the transition temperatures determined by DSC measurements upon heating. The corresponding open symbols were determined upon cooling. Plusses, crosses, and diamonds indicate the α, γ, and γ' phases, respectively, assigned on the basis of the powder X-ray diffraction measurements. Broken curves in the panels denote eye guides for the solidus curves. The present results are in good agreement with those previously report by Többens *et al*. [32].

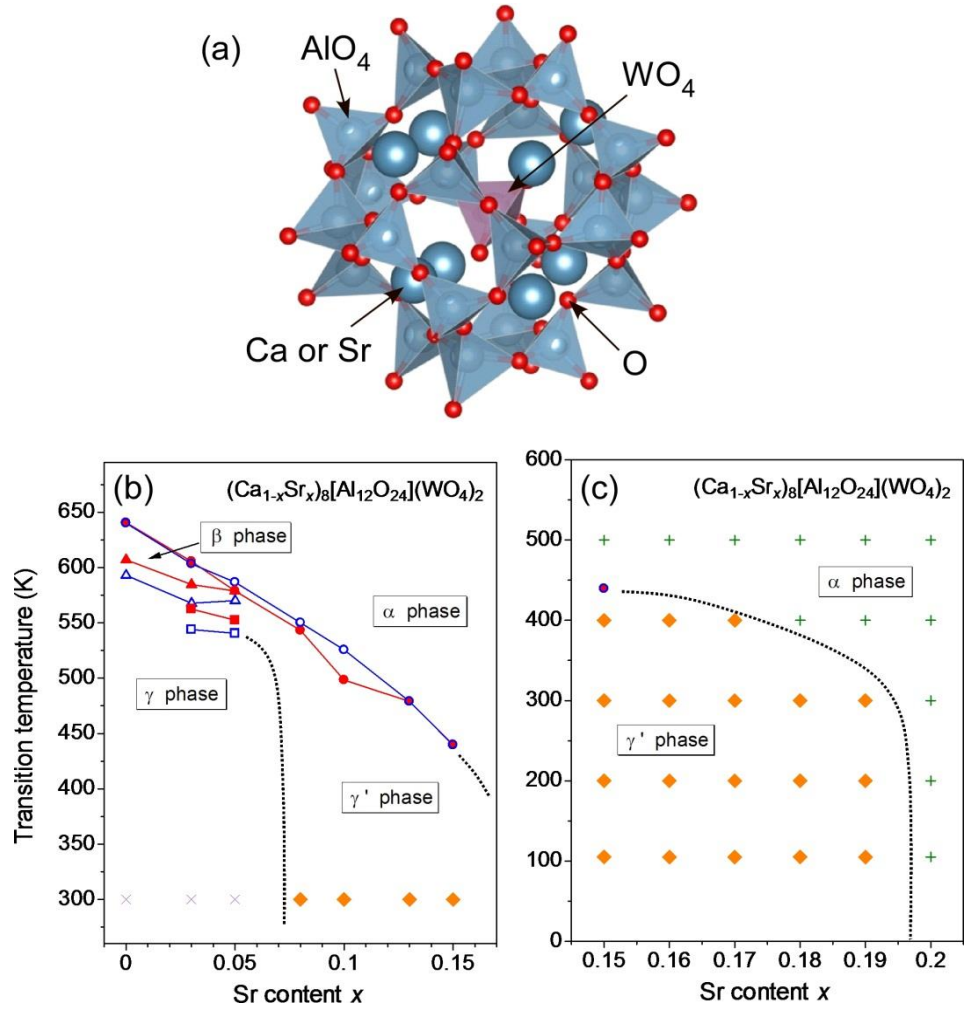

FIG. 1. (a) The crystal structure and the phase diagram of $C_{1-x}S_xAW$ over the ranges (b) $0 \leq x \leq 0.15$ and (c) $0.15 \leq x \leq 0.20$, as determined by DSC and powder X-ray diffraction measurements in the present study.

Figure 2 shows the temperature dependence of dielectric permittivity for $C_{1-x}S_xAW$ ($x = 0$, 0.03, 0.10, 0.16). Squares, circles, triangles, and crosses indicate results for $x = 0$, 0.03, 0.10, and 0.16, respectively. Solid, broken, and chain lines in the figure respectively denote the horizontal axes for $x = 0.03$, 0.10, and 0.16, whose levels were systematically raised for visibility. The arrows in the figure denote several dielectric anomalies indicating the phase transitions in $C_{1-x}S_xAW$. At $x = 0$, a successive phase transition was detected at temperatures above 600 K. The phase-transition temperature decreased upon Sr substitution at $x = 0.03$. With further substitution, $C_{1-x}S_xAW$ underwent a single phase transition, consistent with the phase diagram constructed on the basis of the DSC measurements. Notably, the dielectric permittivity of $C_{1-x}S_xAW$ showed a subtle temperature variation even around the phase-transition temperature and remained below $\varepsilon' \approx 10$ over the entire investigated temperature range in the present study. This behavior contrasts sharply with that for a *proper* ferroelectric phase transition, which generally shows a divergent increase of the dielectric permittivity at temperatures near the phase-transition temperature.

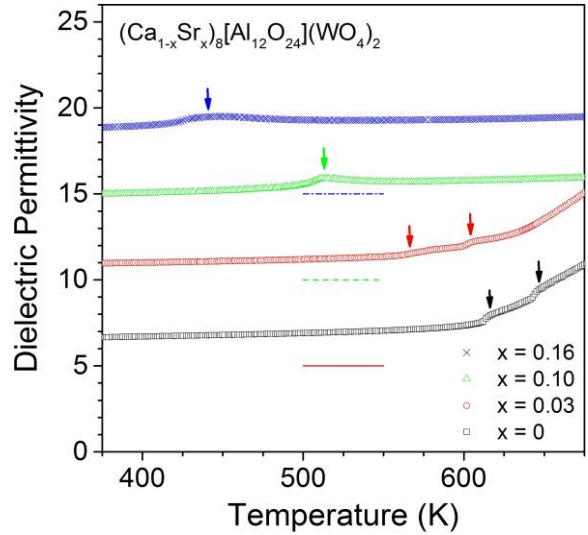

FIG. 2. The temperature dependence of dielectric permittivity for $C_{1-x}S_xAW$. Data points for $x = 0$, 0.03, 0.10, and 0.16 are plotted as squares, circles, triangles, and crosses, respectively. See text for details.

Figure 3 shows the temperature dependence of the spontaneous polarization for $C_{1-x}S_xAW$ ($x = 0.03, 0.10, 0.16$). Circles, triangles, and crosses in the figure denote results for $x = 0.03, 0.10$, and 0.16, respectively. The data for $x = 0$ is not shown because the pyroelectric measurements for $C_{1-x}S_xAW$ at $x = 0$ were unsuccessful as a consequence of its high transition temperature. We observed positive and negative spontaneous polarizations when opposite bias fields were applied during the field-cooling processes in the pyroelectric measurements, indicating switchable spontaneous polarization of $C_{1-x}S_xAW$. In the composition with $x = 0.03$, as denoted by the circles in Fig. 3, a two-step increase in the spontaneous polarization is observed at approximately 600 K and 580 K, corresponding to the successive phase transition. An increase of the Sr-content $x$ suppresses the formation of the intermediate phase between these two transition temperatures, as also clarified by the DSC and dielectric measurements. The value of the spontaneous polarization at room temperature, however, increases with $x$, reaching 0.04 μC/cm$^2$ at $x = 0.16$. This value is much smaller than that of conventional ferroelectric oxides [16]. Such a small spontaneous polarization is often observed in so-called *improper* ferroelectrics, as represented by $Gd_2(MoO_4)_3$ and the multiferroic $TbMnO_3$ [26,28]. The small spontaneous polarization and the weak dielectric anomaly at temperatures near the phase-transition temperature indicate improper ferroelectricity in $C_{1-x}S_xAW$.

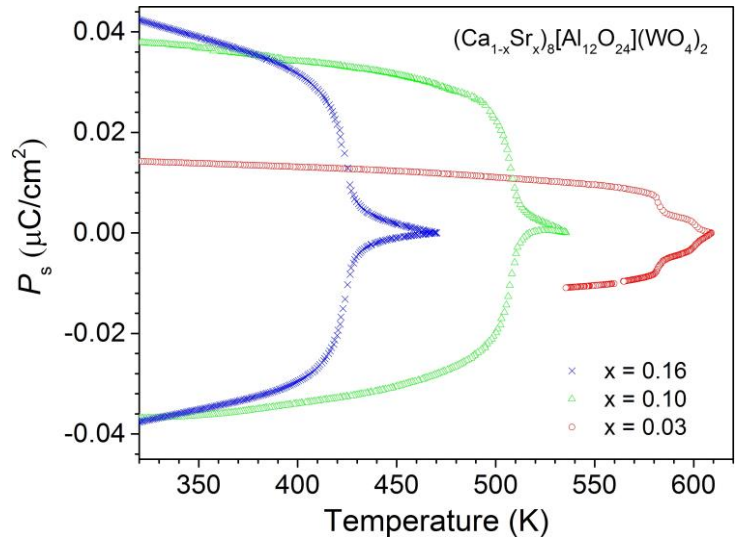

FIG. 3. The temperature dependence of spontaneous polarization for $C_{1-x}S_xAW$. Data points for $x = 0.03$, 0.10, and 0.16 are plotted as circles, triangles, and crosses, respectively. See text for details.

Here, we briefly discuss structural aspects of the phase transition of $C_{1-x}S_xAW$ by analyzing the structural variations of $C_{1-x}S_xAW$ ($x = 0$) through the phase transitions. Figure 4 presents the crystal structures of (a) the highest-symmetry phase, $I\bar{4}3m$, and (b) the lowest-symmetry phase, $Aea2$, observed at 700 K and 300 K, respectively. The squares in the panels denote the unit cells in the phases. The anisotropic thermal vibration is expressed by thermal ellipsoids only for the highest-symmetry phase. In the highest-symmetry phase, as shown in panel (a), each $WO_4$ tetrahedron is in a disordered state among six equivalent orientations in the void of the $[AlO_2]_{12}^{12-}$ framework [22]. The expanded thermal ellipsoids furthermore indicate that potential barriers separating each orientation are smooth, rendering the motion of the $WO_4$ tetrahedra almost rotatory. In synchronization with the disordering of $WO_4$ tetrahedra, the $[AlO_2]_{12}^{12-}$ framework liberates the distortions associated with each orientation of $WO_4$ tetrahedra. $Ca^{2+}$ cations, by contrast, have small thermal ellipsoids, suggesting that their thermal fluctuations are relatively weak. In the lowest-symmetry phase, as shown in panel (b), the $WO_4$ tetrahedra settle at each orientation. The ordering of the $WO_4$ orientation should play an essential role in the phase transition of $C_{1-x}S_xAW$.

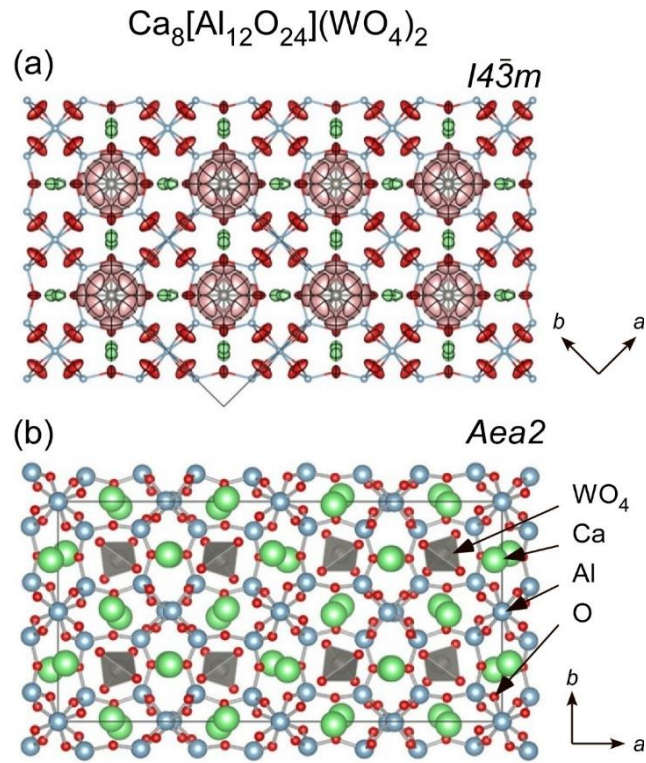

FIG. 4. The crystal structures of $C_{1-x}S_xAW$ in (a) the highest- and (b) the lowest-temperature phases, as determined by structural analyses using powder X-ray diffraction measurements. Solid squares in the panels denote the unit cells for both phases.

To clarify the dynamical aspect of the phase transition in $C_{1-x}S_xAW$, we performed molecular dynamics simulations. Figure 5 shows traces of $WO_4$ motions calculated at (left) 1000 K, (center) 700 K, and (right) 300 K. Yellow dots in the figure indicate variations of positions for apical oxygens in the $WO_4$ tetrahedra with time. As evident in the figure, the trace of the apical oxygens is found to spread around the tungsten cation in the void of the $[AlO_2]_{12}^{12-}$ framework at 1000 K. This spherical distribution of apical oxygens indicates that $WO_4$ tetrahedra rotate almost freely because of the smooth potential barriers at high temperatures. As the temperature decreases to 700 K, as shown in the figure, the spatial distribution of apical oxygens condenses over four positions around the tungsten cation, suggesting ordering of the $WO_4$ orientation. Although the ordering temperature is slightly different from that observed in the experiments, likely because of the limited accuracy of the present calculations, the results of the molecular dynamics simulations clearly demonstrate that the phase transition of $C_{1-x}S_xAW$ is driven by freezing of the nearly free rotation of $WO_4$ tetrahedra in the voids of the $[AlO_2]_{12}^{12-}$ framework. Furthermore, with a further decrease of the temperature to 300 K, the distribution of apical oxygens becomes more localized, indicating that the fluctuation of $WO_4$ weakens in an ordered orientation. Notably, the phase transition to the intermediate phase is not addressed in the present study.

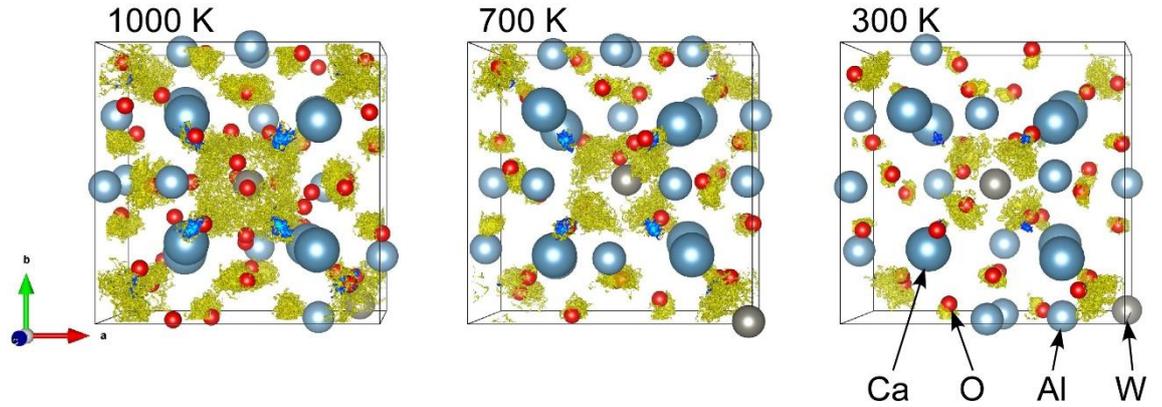

FIG. 5. The fluctuation of $WO_4$ tetrahedra in the void of the $[AlO_2]_{12}^{12-}$ framework at 1000 (left), 700 (center), and 300 K (right), as calculated using molecular dynamics simulations for $C_{1-x}S_xAW$ with $x = 0$. Yellow dots in the figure indicate variations of positions for apical oxygens in the $WO_4$ tetrahedra with time.

To better understand the phase transition of $C_{1-x}S_xAW$, we investigated the lattice dynamics by first-principles calculations. Figure 6 shows the phonon dispersion in $C_{1-x}S_xAW$ at $x = 0$ calculated in the $I\bar{4}3m$ structure at zero temperature, where the area below zero-frequency in the vertical axis indicates imaginary frequency. The results show that several acoustic phonon branches have imaginary frequencies around the Z, X, and N points in the Brillouin zone. Such unstable phonon modes with imaginary frequency are called "soft modes," and their freezing is known to induce structural phase transitions [16]. The displacement patterns of all of these unstable lattice vibrations are observed to have common characteristics involving $WO_4$ tetrahedra vibrations of substantial amplitude. The nearly rotatory fluctuation of $WO_4$ tetrahedra clarified by the molecular dynamics simulations would be composed of these modes, which acquire stable vibrations through thermal fluctuations. As temperature decreases, the frequencies of these modes decrease because of suppression of thermal fluctuations; these modes (or one of them) eventually freeze, triggering the structural phase transition. From this perspective, we can interpret the phase transition of $C_{1-x}S_xAW$ as being soft-mode-driven. Notably, we ignore the imaginary frequency branch between the P and $\Gamma$ points because the present calculations are not sufficiently accurate to discuss a general point in the Brillouin zone.

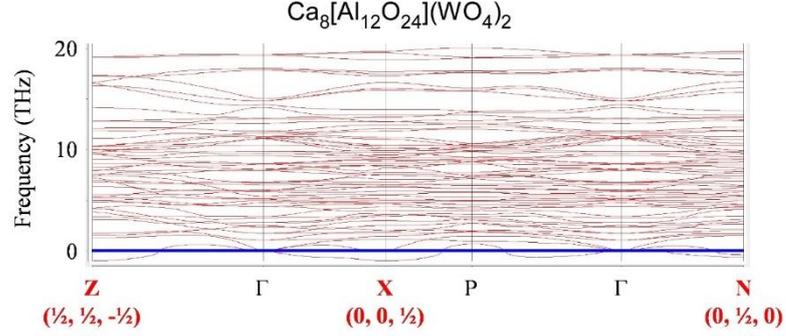

FIG. 6. Phonon dispersion curves for $C_{1-x}S_xAW$ ($x = 0$), as calculated from first principles.

We finally discuss the potential application of $C_{1-x}S_xAW$ for pyroelectric energy harvesting. Two parameters are used to characterize the performance of pyroelectric energy harvester materials: the electromechanical coupling factor and the voltage response. These parameters are defined by $(p^2\theta_h)/(\epsilon'C)$ and $-p/\epsilon'$, respectively, where $p$ is the pyroelectric coefficient at temperature $\theta_h$, $\epsilon'$ is the dielectric permittivity, and $C$ is the heat capacity. Figures 7(a) and (b) show the temperature dependence of the electromechanical coupling factor and the voltage response of $C_{1-x}S_xAW$ with $x = 0.16$ ($C_{0.84}S_{0.16}AW$). Both factors increase as the temperature reaches the phase-transition temperature. Because the dielectric permittivity of $C_{0.84}S_{0.16}AW$ exhibits little temperature dependence, the observed variations of the electromechanical coupling factor and the voltage response mainly originate from the change in the pyroelectric coefficient with changing temperature.

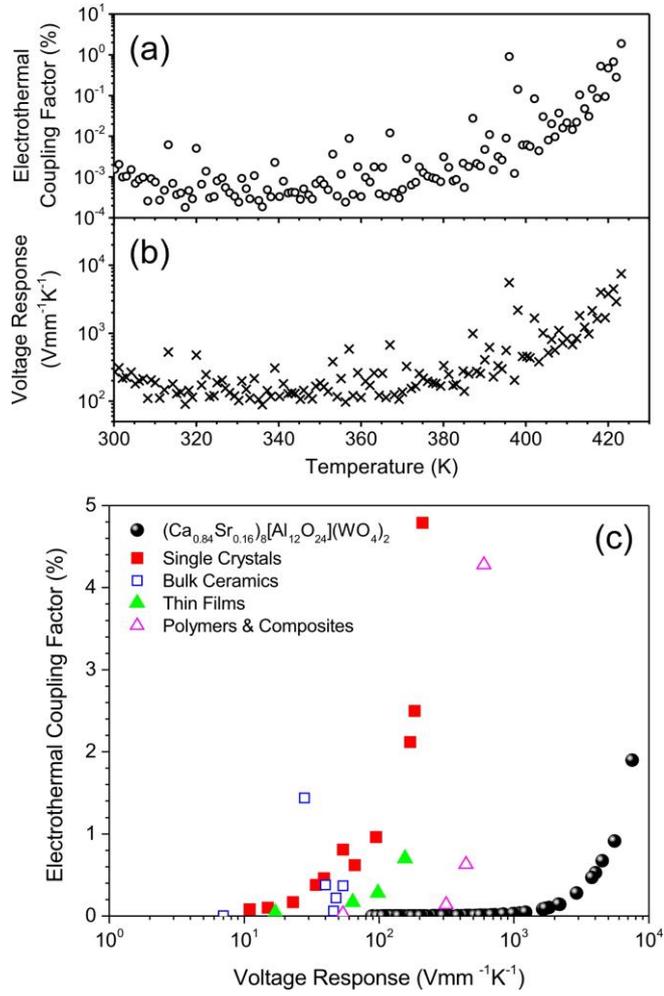

FIG. 7. (a) The electromechanical coupling factor and (b) voltage response of $C_{1-x}S_xAW$ ($x = 0.16$) estimated from the dielectric permittivity and the pyroelectric coefficient measured in the present study. The performance of $C_{1-x}S_xAW$ ($x = 0.16$) as pyroelectric energy harvesters is plotted in panel (c) and is compared with the performance of other materials of several forms. See text for details.

To demonstrate the potential for $C_{0.84}S_{0.16}AW$ as a pyroelectric energy harvester, we mapped its electromechanical coupling factor and voltage response, as shown in Fig. 7(c), along with those for other materials, whose data were extracted from the literature [34]. The results for $C_{0.84}S_{0.16}AW$ are plotted as closed circles in the panel; closed squares, open squares, closed triangles, and open triangles denote the results for single crystals of relaxor PMN-PT, the bulk ceramics, the thin films, and PZT/PVDF-HFP composites, respectively. We note here that data points for $C_{0.84}S_{0.16}AW$ were obtained using the same sample at different temperatures. The most prominent feature of $C_{0.84}S_{0.16}AW$ is its large voltage response that is larger than the voltage responses of the other systems represented in the figure by several orders of magnitude, where a large value of the voltage response indicates a short response time in the device. This large voltage response stems from the unique dielectric response of $C_{0.84}S_{0.16}AW$; because the voltage response is defined by $-p/\epsilon'$, the small dielectric permittivity of the $C_{0.84}S_{0.16}AW$ favors a high value. Furthermore, in the case of $C_{0.84}S_{0.16}AW$, the permittivity exhibits little temperature variation even near the phase-transition temperature, just below which the pyroelectric coefficient $p$ becomes maximum. This behavior dramatically enhances the voltage response near the phase-transition temperature. Such stable dielectric permittivity is specific to improper ferroelectric materials and cannot be attained with proper ferroelectric materials. In the case of the electromechanical coupling factor, $C_{0.84}S_{0.16}AW$ attains a value of 2% at temperatures near the phase-transition temperature. Although this value is approximately one-third of the maximum observed in the case of PMN-PT single crystals, the large voltage response of $C_{0.84}S_{0.16}AW$ imparts this material with strong potential for use as a pyroelectric energy harvester. Enlarging the pyroelectric coefficient of aluminate sodalite is the key to the further development of these materials' improper ferroelectricity.

## IV. CONCLUSION

In summary, we have demonstrated the ferroelectricity of $C_{1-x}S_xAW$ by observing switchable spontaneous polarization via pyroelectric measurements. The small spontaneous polarization and weak dielectric anomaly even at temperatures near the phase-transition temperature indicate the improper nature of the ferroelectricity in $C_{1-x}S_xAW$. Our comprehensive study involving structural analyses, molecular dynamics simulations, and first-principles calculations clarifies that the phase transition of $C_{1-x}S_xAW$ is driven by the freezing of fluctuating $WO_4$ tetrahedra among equivalent orientations in the voids of the $[AlO_2]_{12}^{12-}$ framework. The considerably large voltage response with a large electromechanical coupling factor of $C_{1-x}S_xAW$, which were estimated from the present results, demonstrate the strong potential of stuffed aluminate sodalites as pyroelectric energy harvesters as a result of their unique dielectric properties stemming from their improper ferroelectric phase transition. Because stuffed aluminate sodalites are generally composed of abundant and nontoxic elements, the present study provides a path for the development of functional energy harvesting devices with environmentally friendly oxides.

## ACKNOWLEDGMENTS

This work is partially supported by a Grant-in-Aid for Young Scientists (A) (No.16H06115), MEXT Element Strategy Initiative Project, Ministry of Education, Culture, Sports, Science and Technology of Japan through the Grants-in-Aid for Scientific Research on Innovative Areas "Nano Informatics" (25106008) from JSPS, MEXT Elements Strategy Initiative to Form Core Research Centers, and the "Materials Research by Information Integration" Initiative ($MI^2I$) of the Support Program for Starting Up Innovation Hubs from the Japan Science and Technology Agency (JST). The authors thank Professors E. Nishibori, M. Takata, and M. Sakata for their assistance with the Rietveld analysis. The SR experiments were conducted with the approval of the Japan Synchrotron Radiation Research Institute (JASRI; Proposal Nos. 2014B1468, 2015A1425, and 2015A0074).